\documentclass[11pt]{article}

\pdfoutput=1
\usepackage{jheppub}
\usepackage{diagbox,latexsym,amssymb,amsfonts,amsmath,slashed}
\usepackage{graphicx}
\usepackage{mathrsfs}
\newcommand{\f}[2]{\frac{#1}{#2}}

\newcommand{\la}{\langle}

\newcommand{\ra}{\rangle}

\renewcommand{\Re}{{\rm Re}\,}

\newcommand{\ml}{m}
\newcommand{\Sud}{\Sigma}
\newcommand{\DSt}{\Delta\widetilde\Sud}

\newcommand{\Budapest}{
ELTE E\"otv\"os Lor\'and University, Institute for Theoretical Physics,
P\'azm\'any  P.\ s.\  1/A,  H-1117,  Budapest,  Hungary}
\newcommand{\Frankfurt}{Institute for Theoretical Physics, Goethe Universit\"at Frankfurt, D-60438 Frankfurt am Main, Germany}
\newcommand{\Atomki}{Institute for Nuclear Research of the Hungarian Academy of
Sciences, Bem t\'er 18/c, H-4026 Debrecen,  Hungary}
\newcommand{\Bonn}{HISKP(Theory), University of Bonn, Nussallee 14-16, D-53115 Bonn, Germany}

\author[a]{G.~Endr\H{o}di,}
\author[b]{M.~Giordano,}
\author[b]{S.~D.~Katz,}
\author[b,c]{T.~G.~Kov\'acs,}
\author[d]{and F.~Pittler}

\affiliation[a]{\Frankfurt}
\affiliation[b]{\Budapest}
\affiliation[c]{\Atomki}
\affiliation[d]{\Bonn}

\emailAdd{endrodi@th.physik.uni-frankfurt.de}
\emailAdd{giordano@bodri.elte.hu}
\emailAdd{katz@bodri.elte.hu}
\emailAdd{kgt@atomki.mta.hu}
\emailAdd{pittler@hiskp.uni-bonn.de}

\title{
  Magnetic catalysis and inverse catalysis for heavy pions
  }

\abstract{
We investigate the QCD phase diagram for nonzero background 
magnetic fields using first-principles lattice simulations. 
At the physical point (in terms of quark masses),
the thermodynamics of this system is controlled by two opposing
effects: magnetic catalysis (enhancement of the quark condensate) 
at low temperature and inverse magnetic catalysis (reduction of the
condensate) in the transition region. While the former is known to be
robust and independent of the details of the interactions, inverse
catalysis arises as a result of a delicate competition, effective only
for light quarks. By performing simulations at different quark masses,
we determine the pion mass above which inverse catalysis does not 
take place in the transition region anymore.
Even for pions heavier than this limiting value -- 
where the quark condensate undergoes magnetic catalysis -- 
our results are consistent with the notion that the transition
temperature is reduced by the magnetic field. 
These findings will be useful to guide low-energy models and 
effective theories of QCD.}

\begin{document}

\maketitle

\section{Introduction}
\label{sec:intro}

Strongly interacting matter at finite temperature in the presence of
external magnetic fields has been the subject of intense research in
recent years (see, e.g.,
Refs.~\cite{Fraga:2012rr,Andersen:2014xxa,Miransky:2015ava,Kharzeev:2015znc}
for recent reviews). Besides physical applications in the study of
heavy ion collisions, neutron stars, and the early Universe, this
topic is of considerable interest for a better theoretical
understanding of Quantum Chromodynamics (QCD) in the presence of
external sources.  In this respect, nonperturbative studies by means
of numerical calculations on the lattice have shown a richer variety
of effects than initially expected. Perturbative and model
calculations led one to expect that, regardless of the temperature,
one would find an increase of the quark condensate as the magnitude
$B$ of the magnetic field was increased, a phenomenon called {\it
  magnetic catalysis} (MC)~\cite{Shovkovy:2012zn}, and a corresponding
increase of the (pseudo)critical temperature $T_c$. This was initially
confirmed by lattice studies~\cite{DElia:2010abb}, but the situation
changed as the numerical calculations were made more precise. For
physical quark masses and on fine lattices, it turned out that while
MC is displayed away from the critical region, near $T_c$ the quark
condensate decreases with $B$, i.e., {\it inverse magnetic catalysis}
(IMC) is found, and correspondingly $T_c$
decreases~\cite{Bali:2011qj,Bali:2012zg}. This behavior, originally 
observed for $B<1\textmd{ GeV}^2$, was later found to persist for
stronger magnetic fields and it was argued that $T_c(B)$ monotonically 
decreases up to asymptotically large magnetic
fields~\cite{Endrodi:2015oba}. Results supporting IMC were also 
obtained by further lattice
simulations~\cite{Ilgenfritz:2013ara,Bornyakov:2013eya,Bali:2014kia}.  
It was then believed that MC and IMC corresponded to $T_c$ being
respectively an increasing or decreasing function of $B$, but a recent
study has shown that as the pion mass is increased, the behavior near
$T_c$ crosses over from IMC to MC, while $T_c$ remains a decreasing
function of $B$ all along~\cite{DElia:2018xwo}. 

On the theoretical side, a full understanding of the microscopic
mechanism responsible for these effects is still lacking.
In this respect, it is useful to recall, following
Refs.~\cite{DElia:2011koc,Bruckmann:2013oba}, that the magnetic field
enters the calculation of the condensate both directly through the
observable, and indirectly through the fermion determinant
contributing to the weight of the gauge configurations. The
corresponding effects are called {\it valence effect} and {\it sea
  effect}, respectively. As a matter of fact, the magnetic field has a
catalytic effect on the spectrum of the Dirac operator in a given
gauge configuration, increasing the density of low modes and therefore
the condensate. The valence effect therefore always acts in the
direction of MC: in particular, the valence condensate,
obtained by averaging over configurations with $B=0$ in the fermionic
determinant, increases with $B$ at all temperatures. When reweighting
the valence condensate to the full one, including a nonzero $B$ in the
determinant, those configurations with a larger change in the spectral
density near the origin will be suppressed more: this is the sea
effect, which is expected to act in the direction of IMC. In the end,
it is the balance between the two effects that determines whether MC
or IMC will take place.
Since the magnetic field couples to the gauge field only
indirectly through the fermionic determinant, it is the sea effect
which is responsible for the observed changes in the confining
properties of the theory -- like the Polyakov loop expectation
value~\cite{Bruckmann:2013oba} or the static quark-antiquark
potential~\cite{Bonati:2016kxj}. 

It is clear from the discussion above that there are two main issues
that need to be clarified to fully explain the effect of an external 
magnetic field in QCD. The first issue is the detailed mechanism that
leads to an enhanced density of low modes of the Dirac operator when
$B$ is nonzero. While for free quarks the degeneracy of the Landau
levels is responsible for this enhancement~\cite{Gusynin:1995nb}, 
in strongly interacting QCD these levels are in general not well
defined anymore. Remarkably, the lowest Landau level can still be
meaningfully identified and was shown to quantitatively explain the 
increase of the  quark condensate for strong magnetic fields on the
lattice~\cite{Bruckmann:2017pft}. 

The second issue is the delicate balance between the catalytic
drive of the valence effect and the anticatalytic drive of the sea
effect. This amounts to investigating the circumstances under which
IMC is realized around the transition temperature. This is
particularly relevant for the interpretation of the IMC
phenomenon. Recently, a multitude of low-energy models and effective
theories have been employed to explain the lattice findings about IMC
(see, e.g.,
Refs.~\cite{Kojo:2012js,Fraga:2012fs,Fukushima:2012kc,Fraga:2012ev,Endrodi:2013cs,Chao:2013qpa,Kamikado:2013pya,Ferrer:2014qka,Yu:2014sla,Mueller:2014tea,Fayazbakhsh:2014mca,Rougemont:2015oea,Mamo:2015dea,Dudal:2015wfn,Mao:2016fha,Evans:2016jzo,Gursoy:2016ofp,Pagura:2016pwr,Giataganas:2017koz,Gursoy:2018ydr,Rodrigues:2018pep}).   
In most of these settings magnetic catalysis arises naturally,
but to reproduce inverse catalysis around $T_c$ turned out to require
a tuning of model parameters as functions of the magnetic field (see,
e.g., 
Refs.~\cite{Farias:2014eca,Ferreira:2014kpa,Ayala:2014iba,Ayala:2014gwa}). In 
several cases such a reparameterization only sufficed to achieve a
reduction in $T_c(B)$ for low magnetic fields, whereas for higher $B$
an increasing transition temperature was observed, see for example
Refs.~\cite{Fraga:2013ova,Braun:2014fua,Andersen:2014oaa,Mueller:2015fka}.  
In summary, in recent years the magnetic field-temperature phase
diagram grew out to be a highly non-trivial testing ground for QCD
models. The determination of additional details of this phase diagram
-- like the effect of changing the quark masses -- will therefore
further contribute to a better understanding of the limitations of
such effective descriptions. 

In this paper we will make a step towards a better understanding of
this second issue. Our purpose is to study how the catalytic or
anticatalytic effect of the magnetic field depends on the
pion mass, or equivalently on the mass $m$ of the light quarks,
pinning down the limiting value at which IMC turns into MC. For
each $m$, we do this at the corresponding critical temperature
$T_c= T_c(m)$, and at a fixed value of the magnetic field in
physical units. The dependence on the pion mass was also the subject 
of Ref.~\cite{DElia:2018xwo}. Here we employ a larger set of pion
masses to follow more closely the transition from IMC to MC. Moreover,
we employ a different, mass-independent scale-setting procedure to
assess the robustness of the qualitative picture obtained in
Ref.~\cite{DElia:2018xwo} against different ways to build QCD for
unphysical pion masses.

The plan of the paper is the following. In Section \ref{sec:num_setup} we
give the details of our calculation, including the determination of
$T_c$ and setting of the physical scale. In Section \ref{sec:obs} we
specify our observables. In Section \ref{sec:results} we
discuss our numerical results. Finally, in Section \ref{sec:concl} we
draw our conclusions and show our prospects for the future.

\section{Numerical setup and methods}
\label{sec:num_setup}

We perform our numerical calculations on $N_s^3\times N_t$ lattices
using the tree-level Symanzik improved gauge action with three flavors
of stout improved rooted staggered quarks. We fix the strange quark
mass to its physical value and vary the light quark mass $m=m_{\rm
  ud}$ between its physical value and the strange quark mass, i.e.,
between the physical and the $N_f=3$ flavor symmetric point, using the
values $\ml/\ml_{\rm phys}\in
\{1,\,4,\,8,\,12,\,16,\,18,\,20,\,28.15\}$. The details of our lattice 
ensembles, the line of constant physics and the lattice scale
$a(\beta)$ are described in
Refs.~\cite{Aoki:2005vt,Borsanyi:2010cj,Bali:2011qj}. We adopt a
mass-independent scale-setting scheme, using the results of
Ref.~\cite{Borsanyi:2010cj} for the lattice scale determined at the
physical point. In order to estimate the size of finite-spacing
effects in the scale setting, we have alternatively set the
lattice spacing using the $w_0$ scale~\cite{Borsanyi:2012zs} computed
in the $N_f=3$ system, making use of the continuum value
$w_0=0.153~\mathrm{fm}$~\cite{Borsanyi:2016ksw}. Notice that both
procedures rely on a mass-independent scale setting, 
and are expected to lead to the same continuum results. 
We remark furthermore that the lattice scale could also be set in a
mass-dependent manner -- this approach was followed in
Ref.~\cite{DElia:2018xwo}, which employs $w_0$ at the physical point 
and assumes that it is independent of $m$. Since there is no preferred
choice when dealing with physics off the real world, a comparison
between different scale-setting procedures does not assess a
systematic error, but rather the robustness of the resulting
qualitative pictures. A comparison to the results of
Ref.~\cite{DElia:2018xwo} will be provided below.

\begin{figure}[ht!]
\centering 
\includegraphics[width=.55\textwidth,clip]{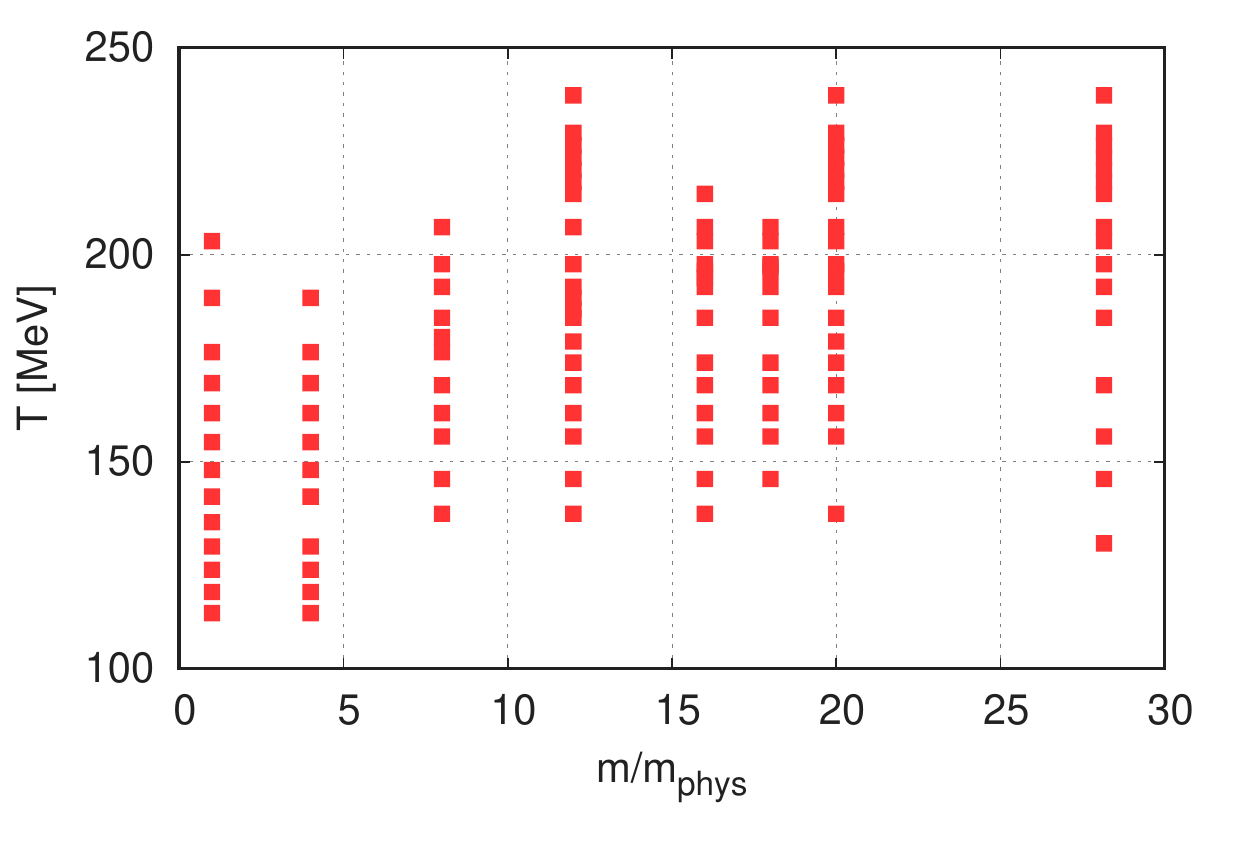} 
\caption{\label{fig:summary_finiteT} The points on the $T-m$
plane used for configuration generation in this work.}
\end{figure}

For our analysis both zero-temperature runs as well as finite-temperature 
simulations were necessary. We generated $T\approx0$ configurations at
$B=0$ using four different values of the gauge coupling $\beta$
summarized in Tab.~\ref{tab:param_zeroT}. These configurations are
used for the determination of the additive renormalization of the 
condensate and of the lattice scale. The finite-temperature
simulations were performed at fixed lattice spatial volume and 
temporal extension  ($N_s=24$ and $N_t=6$). This translates
approximately to lattice spacings between $0.15~\mathrm{fm}$ and
$0.29~\mathrm{fm}$ with our two-level stout improved action. All of
our finite temperature simulation points are summarized on
Fig.~\ref{fig:summary_finiteT}. For each ensemble we generated
$\mathcal{O}\left(200\right)$ well thermalized configurations
separated by 10 HMC trajectories. In the analysis we compute the
statistical error by the bootstrap procedure with 2000 bootstrap
samples. We are performing fully correlated fits when it is
necessary. 

\begin{table}[ht!]
\centering
\begin{tabular}{|ccc|}
\hline
$\beta$ & $N_s$ & $N_t$ \\
\hline
3.450 & 24 & 32\\
3.555 & 24 & 32\\
3.625 & 28 & 40 \\
3.670 & 32 & 48 \\
\hline
\end{tabular}
\caption{\label{tab:param_zeroT} Bare parameters of our $T\approx0$ 
ensembles.}
\end{table}

To fulfill the periodic boundary conditions, we need to use a
quantized magnetic flux $N_b$ in our simulations. The quantization
condition reads 
\begin{align}
\label{eq:quant_cond}
\left(N_sa\right)^2\cdot q_dB = 2\pi N_b,\quad N_b\in\mathbb{Z},\quad
0\leq N_b< N_s^2\,, 
\end{align}
where the smallest of the quark electric charges enters, that of the
down quark $\vert q_d \vert= e/3$, with $e>0$ being the elementary
charge. In order to be able to resolve it, the magnetic field on our
discretized lattice has to be very small in lattice units, i.e.,
$a^2qB\ll 1$, which translates to $N_b/N_s^2\ll 1$ in terms of the
magnetic flux. In this work we use $N_b\in[11,18]$, which results 
in $N_b/N_s^2 < 5\%$, thus in small discretization errors for $B$.

\section{Observables}
\label{sec:obs}

Our central observable is the light quark condensate
$\la\bar\psi\psi\ra=\la\bar u u + \bar d d\ra/2$. Here we follow the
normalization introduced in Ref.~\cite{Bali:2012zg}, 
\begin{align}
\Sud(B,T,\ml)=\frac{2m_{\rm phys}}{M_\pi^2F^2}\left[\la\bar{\psi}\psi\ra_{B,T,m} -
  \la\bar{\psi}\psi\ra_{0,0,m}\right]+1, 
\label{eq:condrenorm}
\end{align}
which contains the physical pion mass ($M_\pi=135~\mathrm{MeV}$) and 
the chiral limit of the pion decay constant ($F=86~\mathrm{MeV}$) at
$B=0$. The so defined combination is free of additive and
multiplicative divergences and is normalized such that it equals unity
for $T=B=0$ and (according to leading-order chiral perturbation
theory) approaches zero for high temperatures. Using
Eq.~(\ref{eq:condrenorm}), the change of the condensate due to the
magnetic field reads 
\begin{equation}
  \label{eq:rencond2}
\Delta \Sud(B,T,\ml) =
\Sud(B,T,\ml)-\Sud(0,T,\ml) =  \f{2\ml_{\rm phys}}{M_\pi^2F^2}\left[
\la\bar\psi\psi\ra_{B,T,\ml} -\la\bar\psi\psi\ra_{0,T,\ml}
\right]\,.
\end{equation}
Note that here we take into account both ($\it a$) the sea effect by
generating configurations at several values of the (quantized)
magnetic flux, and ($\it b$) the valence effect by using the Dirac
operator at $B>0$ in the measurement. 

Magnetic catalysis and inverse catalysis are distinguished by the 
sign of $\Delta\Sud(B,T,m)$. Instead of mapping out the complete
three-dimensional parameter space, in this work we concentrate on a
one-dimensional subspace 
\begin{equation}
\DSt(m)\equiv\Delta\Sud(B_0,T_c(m,B=0),m)\,.
\label{eq:DStdef}
\end{equation}
Thus, we follow the line of pseudo-critical temperatures
$T=T_c\left(\ml,B=0\right)$ on the $T-\ml$ plane. Since at $T_c$ the
system is maximally sensitive to the fermionic determinant, in this
way we expect anticatalytic effects to be at their strongest for each
value of the light quark mass that we simulate. For the magnetic field
we choose $eB_0=0.6~\mathrm{GeV}^2$, which is a typical value where
the IMC phenomenon occurs~\cite{Bali:2012zg}. On $N_t=6$ lattices at
the physical point, the system exhibits IMC, i.e.\ $\DSt(\ml_{\rm
  phys})<0$, see Ref.~\cite{Bali:2012zg}. We will see below that
increasing $\ml$ increases $\DSt$, eventually turning it positive. 
The limiting quark mass $\widetilde{\ml}$ is defined implicitly by
$\DSt(\widetilde{\ml})=0$.\footnote{In general, the set
  $\Delta\Sud(B,T,m)<0$ is a domain in the $T-B$ plane for each value
  of the quark mass. For physical quark masses, $\ml=\ml_{\rm phys}$,
  this domain includes the point $p=[eB=0.6\textmd{ GeV}^2$,
  $T=T_c(\ml)]$. As the quark mass is increased, the domain
  shrinks. According to our definition, $\widetilde{m}$ is the
  limiting mass, where the point $p$ crosses the border of the
  domain. Choosing $p$ differently will change our result slightly but 
  will not affect the emerging picture qualitatively.} 

Besides the quark condensate, we also determined the average Polyakov 
loop, which has already been identified as the most relevant gluonic
observable for the response of QCD matter to a background magnetic
field~\cite{Bruckmann:2013oba}. It is defined as the average product
of time-like links $U_4$ along a closed temporal loop of minimal
length, 
\begin{equation}
P = \frac{1}{V} \left\langle \sum_{\mathbf x} \Re
  \textmd{Tr}\prod_{t=0}^{N_t-1} U_4(\mathbf{x},t) \right\rangle\,. 
\end{equation}
We also consider the ratio 
\begin{equation}
L_R = P(B,T,m) \big/ P(0,T,m)\,,
\label{eq:LRdef}
\end{equation}
in which the multiplicative divergences cancel (since those are
independent of the magnetic field~\cite{Bruckmann:2013oba}).

\section{Results}
\label{sec:results}

\begin{figure}[ht!]
\centering 
\includegraphics[width=.65\textwidth,clip]{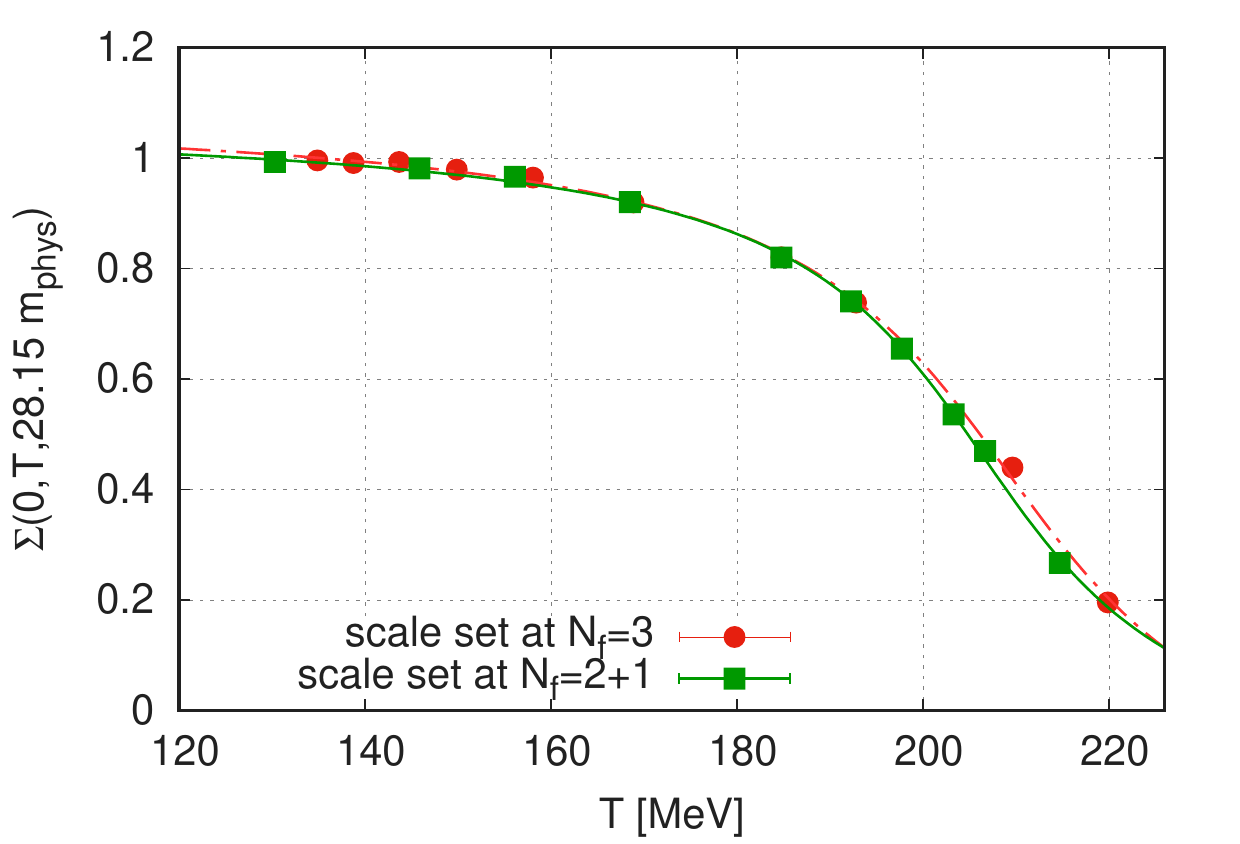}
\caption{Comparison of the two scale setting procedures described in
  the text. Red circles indicate the results using the $w_0$ scale at
  the $N_f=3$ point, while the green squares correspond to the
  $2+1$-flavor LCP~\cite{Borsanyi:2010cj}.}  
\label{fig:scale_comp_w0} 
\end{figure}

To determine $\widetilde{\ml}$, we first performed $B=0$ simulations
to calculate $T_c\left(\ml,B=0\right)$ as a function of $\ml$.  
The pseudo-critical temperature was computed as the inflection point
of $\Sud(0,T,\ml)$, by means of an arctangent fit to the data,
separately for each quark mass represented in
Fig.~\ref{fig:summary_finiteT}. For illustration, we show in
Fig.~\ref{fig:scale_comp_w0} our results for $\Sud$ at the
three-flavor symmetric point with the arctan fits included.  
The two data sets correspond to two independent scale settings: 
$({\it a})$ using $f_K$ at the physical point~\cite{Borsanyi:2010cj},
$({\it b})$ using $w_0$ at the $N_f=3$ point. From the figure it is
apparent that the uncertainty coming from the scale setting is tiny. 
This is also reflected by the extracted inflection points, which agree 
with each other within one standard deviation. At the physical point
we find on our $N_t=6$ lattices that $T_c\left(\ml_{\rm
    phys},B=0\right)=149.9(9)\,{\rm MeV}$, only a few percent away
from the continuum limit $T_c^{\rm cont}=157(4)\,{\rm
  MeV}$~\cite{Aoki:2009sc}. We take this as an indication of small 
finite-spacing effects.

From the $T_c(\ml)$ data we can determine the complete pseudo-critical  
trajectory using an interpolation in the quark mass. We have tried
fits with several functional forms, namely the ``rational'' function  
\begin{equation}
T_c(\ml)=T_c(0)\frac{1+a\,\ml^{c}}{1+b\,\ml^{c}}\,,
\label{eq:chiral_fit}
\end{equation}
and the power-law behavior 
\begin{equation}
T_c(\ml)=T_c(0)+am^{u}(1+ bm^2 + cm^4)\,,
\label{eq:chiral_fit2}
\end{equation}
where we set $u=\frac{1}{\beta\delta}$ with $\beta$ and $\delta$ the
critical exponents of the O(4) or the O(2) universality classes
(see, e.g., Ref.~\cite{Engels:2001bq}). We plot our results in
Fig.~\ref{fig:chiral_fit} against the respective pion masses (for
their determination, see below), and list our resulting fit parameters
in Tables \ref{tab:param_finiteT} and \ref{tab:param_finiteT2}. The
errors of $T_c(m)$ used in the fits include the statistical error and
the systematic error related to the choice of fitting range in the
determination of the inflection point. In the plot a further 2\%
uncertainty due to the determination of the physical scale
(see Ref.~\cite{Aoki:2009sc}) is also included. 

In Fig.~\ref{fig:chiral_fit} we include also the results of the Pisa 
group (Fig.~6 of Ref.~\cite{DElia:2018xwo}) for comparison. While the
different scale-setting procedure obviously leads to quantitatively
different results from ours, the qualitative behaviors match nicely.

As an interesting side result, we determine the critical temperature
in the chiral limit as $T_c(\ml=0)=138(4)~\mathrm{MeV}$. The central
value is obtained averaging the three best fits in Tables
\ref{tab:param_finiteT} and \ref{tab:param_finiteT2}, while the error
is obtained by averaging in quadrature the corresponding statistical
errors and the deviation of the three central values from the mean.

\begin{table}[tbp]
\centering
\begin{tabular}{|r|c|}
\hline & ``rational''  \\
\hline
$T_c(0)~{\rm [MeV]}$ &  140(4) \\
$a$ &  0.17(4) \\
$b$   &  0.09(1) \\
$c$   &  0.8(2) \\ \hline
$\chi^2/{\rm d.o.f.} $ & 0.6 \\
\hline
\end{tabular}
\caption{The parameters of a fit to the $T_c(\ml)$ data 
using the function of Eq.~\protect\eqref{eq:chiral_fit}.}
\label{tab:param_finiteT}
\end{table}

\begin{table}[tbp]
\centering
\begin{tabular}{|r|cc|}
\hline  &  O(4) & O(2)  \\
\hline
 $T_c(0)~{\rm [MeV]}$  & 138(1)               &  135(1)       \\
$a~{\rm[MeV]}$ & 12.2(5)              &  14.5(6)             \\
$b$  & $-6(1)\cdot 10^{-4}$ &  $-4(1)\cdot 10^{-4}$\\
$c$  & $4(1)\cdot 10^{-7} $ &  $3(1)\cdot 10^{-7}$  \\ \hline
 $\chi^2/{\rm d.o.f.} $ & 0.4 & 0.4 \\
\hline
\end{tabular}
\caption{The parameters of a fit to the $T_c(\ml)$ data 
using the function of Eq.~\protect\eqref{eq:chiral_fit2}, with $u$ 
chosen according to the indicated critical behavior.}
\label{tab:param_finiteT2}
\end{table}
  
\begin{figure}[tbp]
\centering 
\includegraphics[width=.6\textwidth,clip]{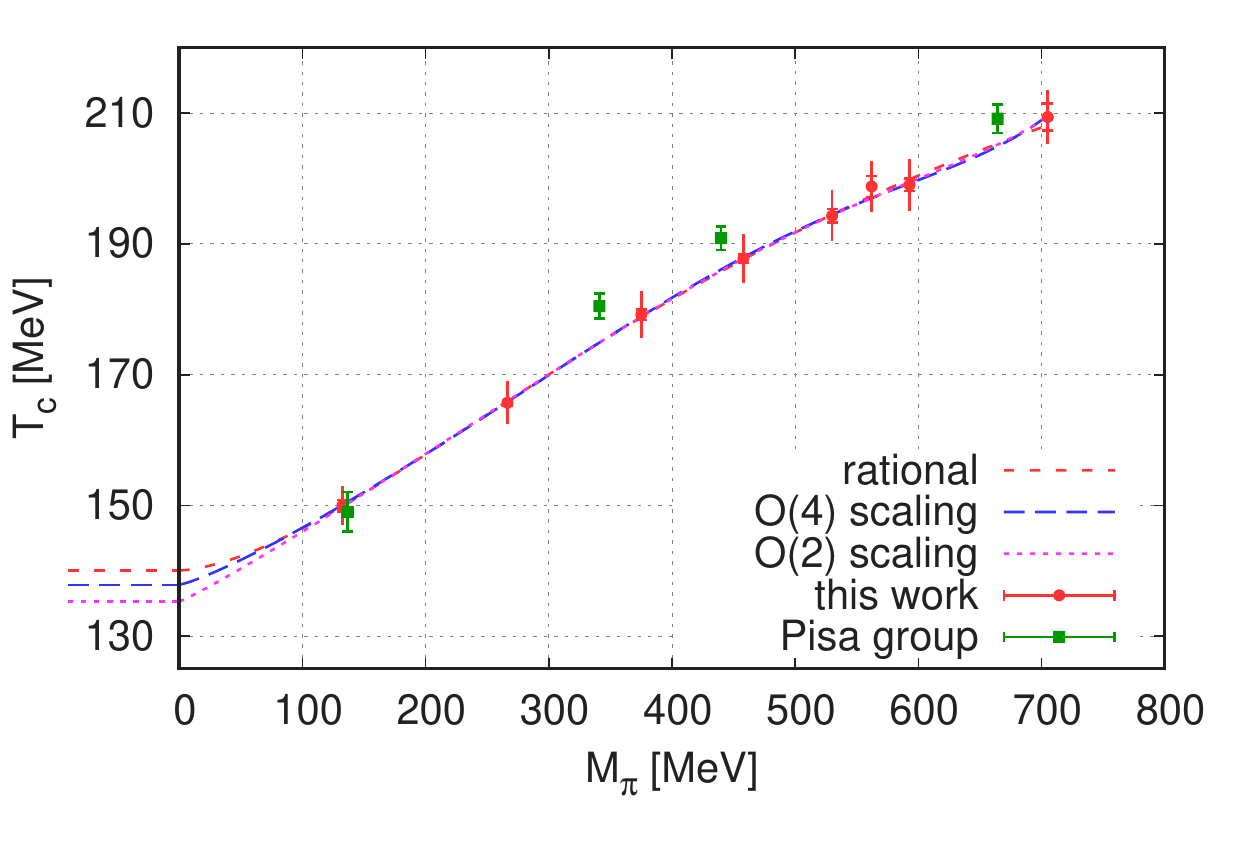}
\caption{\label{fig:pseudocritical} The pseudo-critical trajectory,
  together with the results of fits performed according to
  Eqs.~\eqref{eq:chiral_fit} and \eqref{eq:chiral_fit2}. The extended
  error bars include the systematic effect due to scale setting. Data
  from Ref.~\cite{DElia:2018xwo} are also included for comparison.} 
\label{fig:chiral_fit}
\end{figure}

\begin{figure}[tbp]
\centering 
\includegraphics[width=.6\textwidth,clip]{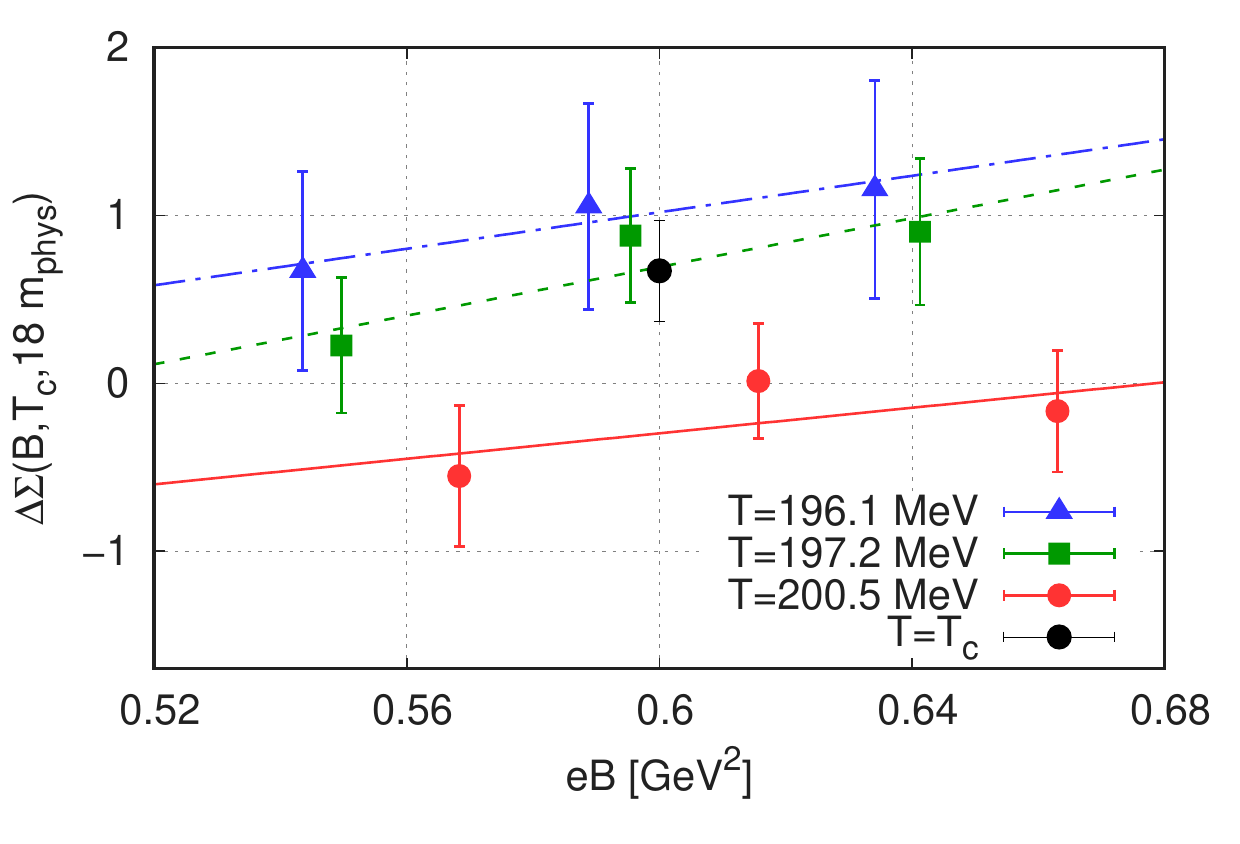}
\caption{\label{fig:scale_comp} 
Visualization of our interpolation scheme for $\ml/\ml_{\rm
  phys}=18$. The pseudo-critical temperature is $T_c=197.3(5)\textmd{
  MeV}$. The lines indicate the results of the interpolation in $B$ to
$eB_0=0.6~{\rm GeV}^2$ at a fixed temperature, and the black filled
circle represents the result  of the final interpolation in the
temperature.}
\label{fig:magn_interp}
\end{figure}

Due to the quantization condition, Eq.~\eqref{eq:quant_cond}, we are
not able to perform simulations at the same physical magnetic field on
all points of the pseudo-critical trajectory. In order to correct for
this, we perform for each quark mass several simulations near
$eB_0=0.6\textmd{ GeV}^2$ and $T_c(\ml)$ and interpolate linearly in
the magnetic field and in the temperature. For illustration we show
our results for the interpolation at a particular light quark mass on
Fig.~\ref{fig:magn_interp}. 

\begin{figure}[tbp]
\centering 
\includegraphics[width=.6\textwidth,clip]{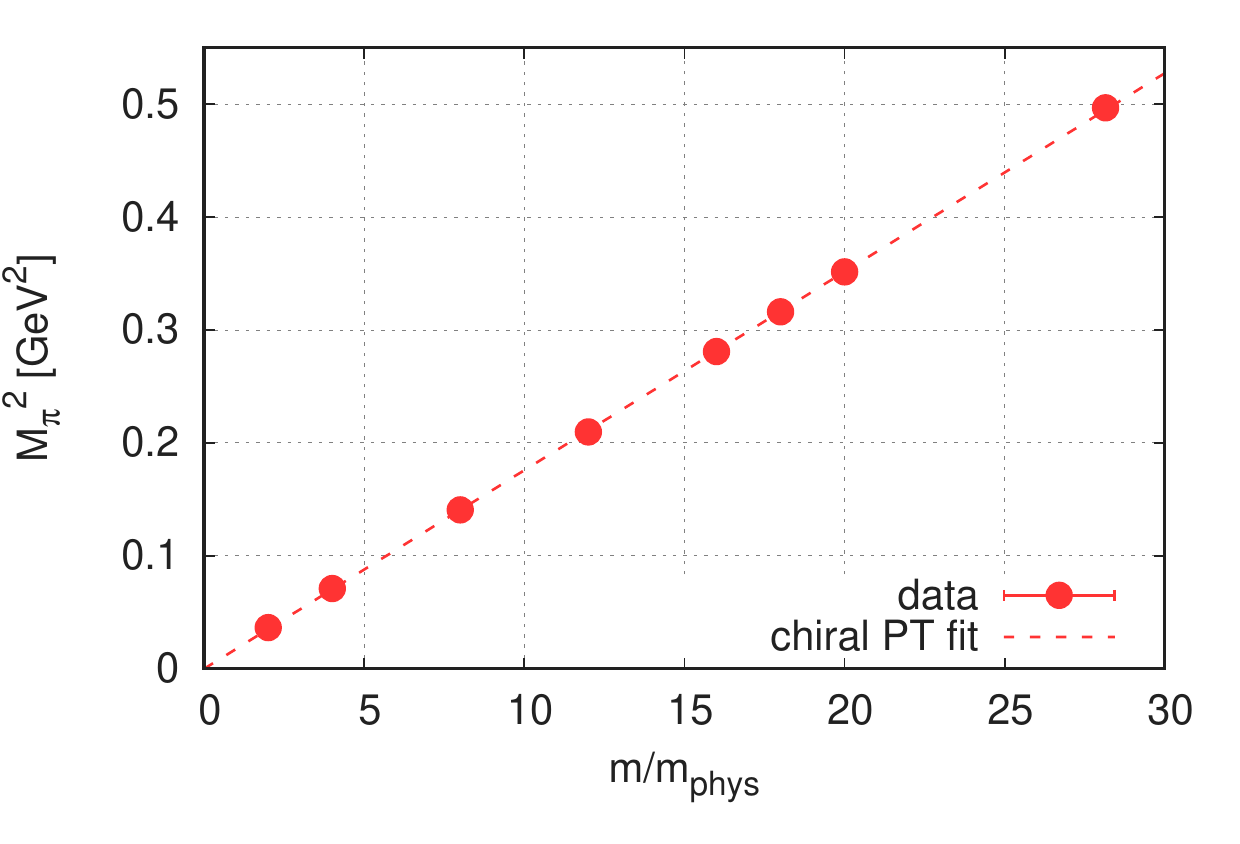}
\caption{The squared pion mass as a 
function of the light quark mass. The dashed line indicates the linear 
dependence predicted by chiral perturbation theory.}
\label{fig:pionmass}
\end{figure}

In order to interpret our results in terms of physical parameters, we
also determined the pion mass $M_\pi$ using our zero temperature
ensembles for several quark masses. We show our results in
Fig.~\ref{fig:pionmass}, which agree well with the prediction of
chiral perturbation theory for the pion mass. We find $M_\pi^2 =
M_0^2\cdot (\ml/m_{\rm phys})$ with $M_0= 132.62(4)~{\rm MeV}$, in
good agreement with the physical pion mass. This dependence is used 
to interpolate the pion mass for intermediate values of $\ml$.

\begin{figure}[tbp]
\centering 
\includegraphics[width=.6\textwidth,clip]{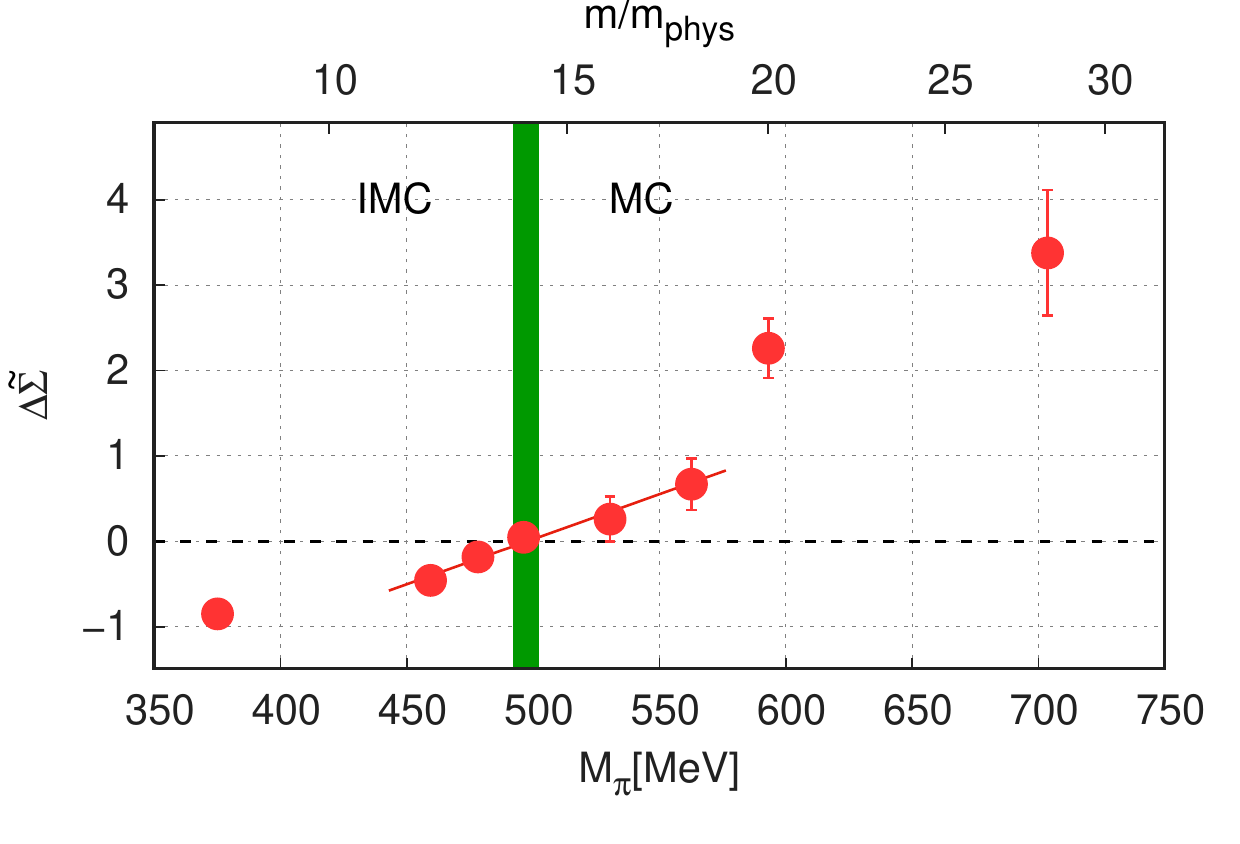}
\caption{The change in the condensate due to the magnetic field
  $eB_0=0.6\textmd{ GeV}^2$ along the pseudo-critical trajectory 
in terms of the pion mass. The green vertical line indicates the limiting pion 
mass, which separates the IMC and MC regions.}
\label{fig:mrat}
\end{figure}

The main result of this paper is shown in Fig.~\ref{fig:mrat}, where
we plot the change of the renormalized chiral condensate $\DSt$ of
Eq.~\eqref{eq:DStdef} against the quark mass (and, equivalently,
against the pion mass). Remember that this quantity measures the
change in the condensate when switching on a magnetic field of
magnitude $eB_0=0.6~{\rm GeV}^2$ at the pseudo-critical temperature
$T_c(\ml)$. The sign of $\DSt$ changes from negative to positive at
the limiting quark mass $\widetilde{\ml}=14.07(55)\,\ml_{\rm phys}$.
The corresponding pion mass equals
$\widetilde{M}_\pi=497(4)~\mathrm{MeV}$. We obtain this value using a
linear interpolation in the interval $M_\pi\in[450,570]~\mathrm{MeV}$
with reduced $\chi^2\simeq 1$. These results show that a change from
IMC to MC in the response of strongly interacting matter to a
background magnetic field takes place for sufficiently heavy pions, as
already observed in Ref.~\cite{DElia:2018xwo}, and allow us to
quantify how heavy pions have to be.  

\begin{figure}[tbp]
\centering 
\includegraphics[width=.6\textwidth,clip]{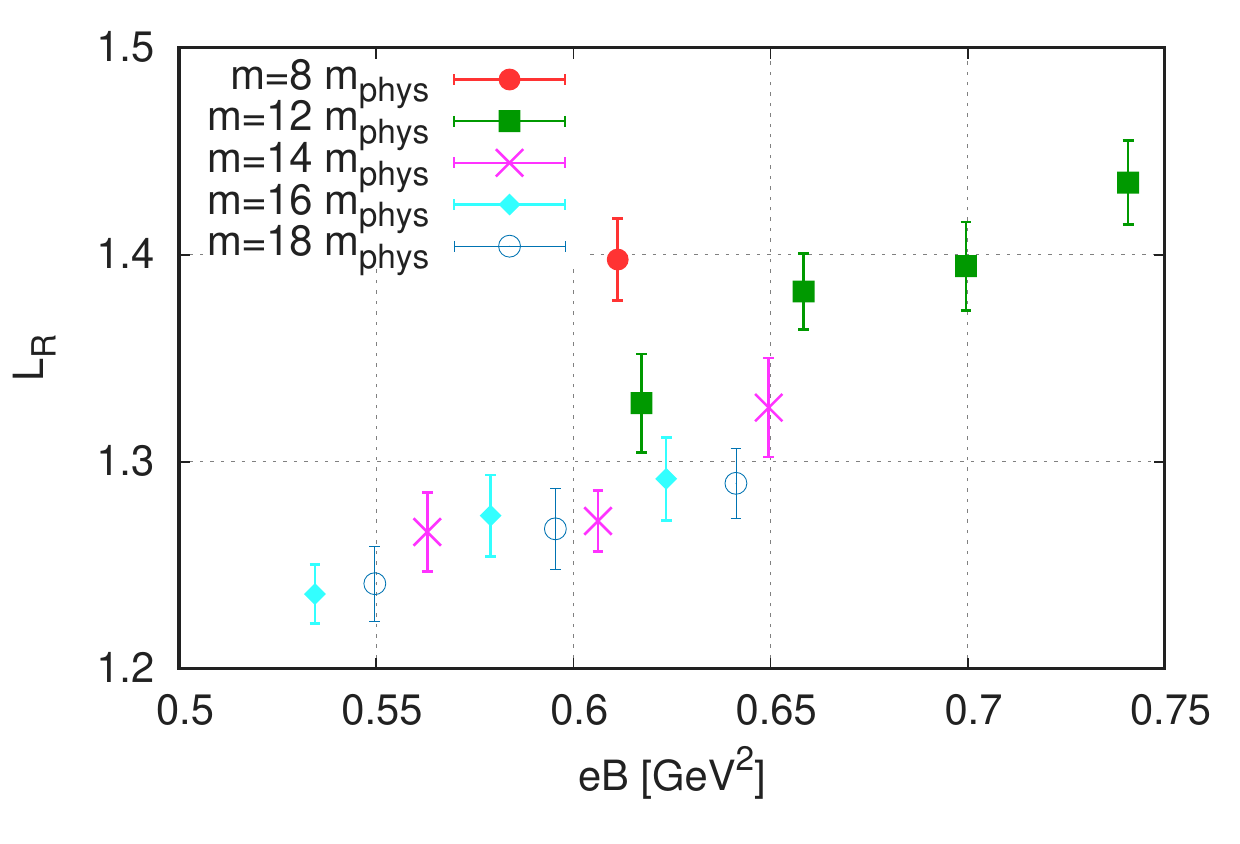}
\caption{The Polyakov loop ratio~\protect\eqref{eq:LRdef} around the point
where inverse magnetic catalysis turns into magnetic catalysis.} 
\label{fig:polyakov_loop}
\end{figure}

Finally, in Fig.~\ref{fig:polyakov_loop} we show the Polyakov loop
ratio~\eqref{eq:LRdef}, which clearly shows that the renormalized
Polyakov loop increases monotonically for all quark masses in the
transition region. This finding is in line with the recent results of
Ref.~\cite{DElia:2018xwo}, where also the inflection point of $P$ was
determined and $T_c(B)$ was shown  to be a decreasing function of $B$
for pion masses up to $M_\pi\approx 660\textmd{ MeV}$, independently
of whether MC or IMC takes place.\footnote{The Polyakov loop is
  expected to be independent of the magnetic field both for
  sufficiently low and for sufficiently high temperatures. Thus, the
  implicit condition $P(T_c)=\textmd{const.}$ is a feasible
  alternative definition for the transition  temperature $T_c$. An
  increase in $P(B)$ for all temperatures therefore results in a
  decreasing $T_c(B)$. Note that an analogous construction does not
  capture the behavior of the quark condensate, since $\Sud$ depends
  on $B$ even at $T=0$.}  
This can perhaps be understood in terms of the inverse correlation
between the Polyakov loop and the reweighting factor of a gauge
configuration due to switching on a magnetic field, observed in
Ref.~\cite{Bruckmann:2013oba}. Such a correlation implies that
configurations with larger values of the average Polyakov loop are
favored in the presence of a magnetic field, compared to the typical
configurations at $B=0$. This pushes the system towards the ordered
phase, thus anticipating the transition and lowering the
pseudocritical temperature for $B\neq 0$. Whether this leads to MC or
IMC at $T_c$ depends instead on the correlation between the chiral
condensate and the reweighting factor. The results of
Ref.~\cite{DElia:2018xwo} and of this paper suggest that while there
is always inverse correlation between the Polyakov loop and the
reweighting factor, the sign of the correlation between the chiral
condensate and the reweighting factor depends on the quark mass.

\section{Conclusions}
\label{sec:concl}

In this paper we have determined the limiting light quark mass (or
equivalently the pion mass) above which QCD does not exhibit inverse 
magnetic catalysis anymore in the transition region. Specifically, 
we considered a fixed magnetic field $eB_0=0.6\textmd{ GeV}^2$ and
evaluated the quark condensate along the pseudo-critical temperature
trajectory $T_c(m,B=0)$. This choice was made so that the sea effect
due to the fermion determinant was as strong as possible.
Our results agree with the general findings of the Pisa group,
reported in Ref.~\cite{DElia:2018xwo}, namely that IMC turns into MC
for large enough pion masses, and also allow to pinpoint the
particular value at which this happens for our choice of magnetic
field. In our setting, we found that the system turns from displaying
IMC to MC at $M_\pi\approx3.7\cdot M_{\pi,\rm phys}$. This value is
consistent with the results of Ref.~\cite{DElia:2018xwo}, although a
quantitative comparison would require to take into account their use
of a different scale-setting procedure. Our results are also
consistent with the preliminary results reported in
Ref.~\cite{Tomiya:2019nym}. The results of this paper were obtained at
a single lattice spacing, but we found indications that finite-spacing
effects are small. While an extrapolation to the continuum is expected
to give (slightly) different quantitative results, we believe that the
qualitative picture of a change from MC to IMC is robust. 
Being based on a first-principles calculation on the lattice,
our results provide a nontrivial testing ground for effective models
aiming at the description of the inverse magnetic catalysis
phenomenon. 

\section*{Acknowledgements}

The authors are grateful to Falk Bruckmann for discussions at the
early stages of this work. This work was partly supported by the DFG 
(Emmy Noether Programme EN 1064/2-1), the Hungarian National Research,
Development and Innovation Office - NKFIH grant KKP126769 and by OTKA
under the grant OTKA-K-113034.

\bibliographystyle{jhep_new}
\bibliography{references}
\end{document}